\documentclass[11pt,english]{article}
\usepackage[T1]{fontenc}
\usepackage[latin9]{inputenc}
\usepackage{geometry}
\usepackage{amsmath}
\usepackage{amssymb}

\makeatletter
\numberwithin{equation}{section}

\date{} 
\usepackage[T1]{fontenc} 
\usepackage{pslatex} 
\usepackage{bbold} 
\usepackage{dsfont} 

\makeatother

\usepackage{babel}
\begin{document}
\setcounter{page}{0}

$\,\,$\vspace{1cm}

\begin{center}
\textbf{\LARGE{}New $\mathbb{Z}_{3}$ strings }
\par\end{center}{\LARGE \par}

\vspace{1cm}

\begin{center}
\textbf{Marco A.C. Kneipp}\footnote{marco.kneipp@ufsc.br}\textbf{
and Paulo J. Liebgott}\footnote{paulo.liebgott@ufsc.br}
\par\end{center}

\begin{center}{\em Departamento de F\'\i sica,\\ 

Universidade Federal de Santa Catarina (UFSC),\\

Campus Universit\'ario, Trindade,\\

88040-900, Florian\'opols, Brazil.  \\ }

\end{center}

\vspace{0.3cm}
\begin{abstract}
We consider a Yang-Mills-Higgs theory with the gauge group $SU(3)$
broken to its center $\mathbb{Z}_{3}$ by two scalar fields in the
adjoint representation and obtain new $\mathbb{Z}_{3}$ strings asymptotic
configurations with the gauge field and magnetic field in the direction
of the step operators. 

\vfill 

\thispagestyle{empty}
\end{abstract}
\newpage

\section{Introduction}

String or vortex solutions may appear naturally in non-Abelian theories
if the vacuum manifold has a nontrivial fundamental group, i.e., $\pi_{1}(G/G_{0})\neq\mathds{1}$,
where $G$ and $G_{0}$ are, respectively, the gauge and the unbroken
gauge groups. These configuration are topological and may be relevant
in many areas such as condensed matter physics \cite{Abrikosov:1956sx},
astrophysics and cosmology \cite{Vilenkin:1981iu} and high energy
physics, in special in Grand Unified Theories \cite{Kibble:1982ae}
and the quark confinement problem \cite{tHooft1975,Mandelstam:1974pi}.
In order to have finite energy per unit length, or finite tension,
the string solutions are asymptotically constructed by gauge transformations
of the vacuum configuration and the associated group elements can
be related to the fundamental group elements. This relation classify
the solutions into topological classes, the strings belonging to the
same equivalence class being homotopic to each other. These features
can be seen in the Abelian Higgs Model \cite{Nielsen:1973cs} which
couples a $U(1)$ gauge field to a complex scalar. This symmetry is
spontaneously broken by a non vanishing vacuum configuration $\phi_{vac}$,
which vanishes the potential. Hence the asymptotic scalar field can
be written as $\phi(\varphi)=\exp(ia(\varphi))\phi_{vac}$ and therefore
provides a mapping from the circle $S^{1}$ at spatial infinity to
the circle $S^{1}$ in the internal space. These mappings can be classified
into homotopy classes and this prevents non trivial solutions to be
continuously deformed into the vacuum configuration. These string
solutions may also appear in theories with non-Abelian groups and
they were initially studied in \cite{deVega:1976rt,deVega:1986eu,deVega:1986hm}. 

It is believed that confinement in the strong coupling regime of QCD
is due to chromoelectric strings called QCD strings. Many properties
of QCD strings have been studied using lattice QCD. On the other hand,
it is believed that QCD string in the strong coupling may be dual
to chromomagnetic strings in weak coupling \cite{tHooft1975,Mandelstam:1974pi}.
Since QCD strings appear in gauge theories with non-Abelian groups
(without $U(1)$ factors) and are believed to be associated to center
elements, in the last years we have analyzed some properties of chromomagnetic
$\mathbb{Z}_{N}$ strings which appear in Yang-Mills-Higgs theories
with arbitrary simple gauge group $G$ broken to its center $\mathcal{Z}(G)$
by two scalar fields \cite{Kneipp:2003ue,Kneipp:2007fg,Kneipp:2008dc}.
These $\mathbb{Z}_{N}$ strings are associated to coweights of $G$
and their topological sectors associated to center elements as was
analyzed in detail in \cite{Kneipp:2007fg}. Similarly to QCD strings,
the tensions of the BPS $\mathbb{Z}_{N}$ strings can satisfy the
Casimir law\cite{Kneipp:2003ue,Kneipp:2007fg}. We showed also that
the magnetic charges of the monopoles, which appear in the first symmetry
breaking, are always integer linear combinations of the magnetic charges
of the $\mathbb{Z}_{N}$ strings, which allows the monopole confinement
by the $\mathbb{Z}_{N}$ strings.

In all our previous works we consider $\mathbb{Z}_{N}$ strings with
gauge field and magnetic field in the direction of the Cartan subalgebra
(CSA). However it is possible to have string or vortex solutions with
gauge fields as combinations of step operators as has been done for
strings in theories with gauge group $SO(10)$ broken to $SU(5)\times\mathbb{Z}_{2}$
\cite{Aryal:1987sn}, $SU(2)$ broken to $\mathbb{Z}_{2}$\cite{Hindmarsh:1994re}
and $SU(2)\times U(1)$\cite{Kibble:1996rh}. In this paper we consider
a Yang-Mills-Higgs theory with gauge group $SU(3)$ broken to its
center $\mathbb{Z}_{3}$ and construct the asymptotic configurations
with gauge fields and magnetic fields as a combination of the $\mathfrak{su}(3)$
step operators which we shall call E-strings.

The paper is organized as follows: in this paper, in section 2 we
define our conventions adopted through the paper and define a new
basis for $\mathfrak{su}(3)$. In section 3 we choose a vacuum configuration
that spontaneously breaks the symmetry and allows for the existence
of $\mathbb{Z}_{3}$ strings. In section 4 we define group elements
associated to generators which are not in the Cartan subalgebra, show
them to satisfy the condition \eqref{eq:intro2} and then obtain the
asymptotic form of the vector and the scalar fields.

\section{Conventions}

Let us consider a Yang-Mills theory with the Lagrangian
\begin{equation}
\mathcal{L}=-\frac{1}{4}G_{\mu\nu}^{a}G^{a\mu\nu}+\frac{1}{2}(D_{\mu}\phi_{s})_{a}^{\ast}(D^{\mu}\phi_{s})_{a}+V(\phi_{s}),\label{eq:strings1}
\end{equation}
where $\phi_{s}$, $s=1,2$, are complex scalar fields in the adjoint
representation of the gauge group $G$ whose Lie algebra is $\mathfrak{g}$.
The covariant derivative is defined as
\[
D_{\mu}\phi=\partial_{\mu}\phi+ie[W_{\mu},\phi],
\]
and the field strength tensor is 
\[
G_{\mu\nu}=\partial_{\mu}W_{\nu}-\partial_{\nu}W_{\mu}+ie[W_{\mu},W_{\nu}].
\]

The so-called Cartan-Weyl basis decomposes the generators of a simple
Lie algebra into the Cartan elements $H_{i}$ and the step operators
$E_{\alpha}$ satisfying
\[
[H_{i},H_{j}]=0,
\]
\[
[H_{i},E_{\alpha}]=\alpha(H_{j})E_{\alpha},
\]
where $\alpha$ is said to be a root of the algebra. Given an algebra
of rank $r$, the roots belong to an $r$-dimensional vector space,
$\Phi(\mathfrak{g})$, called root space. This space is dual to the
space containing the weights $\lambda$ which satisfy $H_{i}|\lambda_{a}\rangle=\lambda_{a}(H_{i})|\lambda_{a}\rangle$.
The basis of these spaces are the simple roots $\alpha_{i}$ and the
fundamental weights $\lambda_{i}$, respectively. The fundamental
co-weights and simple co-roots are defined as $\lambda_{i}^{\vee}=2\lambda_{i}/\alpha_{i}^{2}$
and $\alpha_{i}^{\vee}=2\alpha_{i}/\alpha_{i}^{2}$, respectively,
and they satisfy $\lambda_{i}^{\vee}\cdot\alpha_{j}=\lambda_{i}\cdot\alpha_{j}^{\vee}=\delta_{ij}$.
The step operators have the following commutation relations among
themselves,
\[
[E_{\alpha},E_{-\alpha}]=\begin{cases}
\frac{2\alpha}{\alpha^{2}}\cdot H, & \mathrm{if\ }\alpha=-\beta,\\
N_{\alpha\beta}E_{\alpha+\beta,} & \mathrm{if\ }\alpha+\beta\in\Phi(\mathfrak{g}),\\
0, & \mathrm{if\ }\alpha+\beta\notin\Phi(\mathfrak{g}),
\end{cases}
\]
where $N_{\alpha\beta}$ are antisymmetric coefficients. 

We define the generators 
\[
T_{3}\equiv\sum_{i=1}^{r}\lambda_{i}^{\vee}\cdot H,\quad T_{\pm}=\sum_{i=1}^{r}\sqrt{2c_{i}}E_{\pm\alpha_{i}},
\]
where $c_{i}=\sum_{j=1}^{r}(K^{-1})_{ij}$, with $K_{ij}=2\alpha_{i}\cdot\alpha_{j}^{\vee}$
being the elements of the Cartan matrix associated to $\mathfrak{g}$.
These generators form an $\mathfrak{su}(2)$ algebra
\[
[T_{3},T_{\pm}]=\pm T_{\pm},\quad[T_{+},T_{-}]=2T_{3},
\]
 which is called the principal $\mathfrak{su}(2)$ subalgebra of $\mathfrak{g}$. 

From the so-called principal element,
\[
S=\exp\left(\frac{2\pi iT_{3}}{h}\right),
\]
where $h$ is the Coxeter number of $\mathfrak{g}$, one can show
that the generators of the algebra satisfy
\[
ST^{(n)}S^{-1}=\exp\left(\frac{2\pi in}{h}\right)T^{(n)},\quad n=0,1,\ldots,h-1,
\]
which provides $\mathfrak{g}$ with a $\mathbb{Z}_{h}$ grading,
\[
\mathfrak{g}=g_{0}\oplus g_{1}\oplus\ldots\oplus g_{h-1}.
\]
The elements of the subspace $g_{k}$ has degree $k$ and only $g_{0}$
forms a subalgebra, which corresponds to the Cartan subalgebra formed
by the generators $H_{i}$. Let $\psi$ denote the highest root. We
can expand $\psi$ in the basis of simple roots $\alpha_{i}$ as $\psi=\sum_{i=1}^{r}n_{i}\alpha_{i}$,
where $n_{i}$ are positive integers. Then, the Coxeter number can
be written as $h=1+\sum_{i=1}^{r}n_{i}$ and we can conclude that
\[
\left\{ E_{\alpha_{1}},E_{\alpha_{2}},\,...\,,E_{\alpha_{n}},E_{-\psi},\right\} \subset g_{1},
\]
\[
\left\{ E_{-\alpha_{1}},E_{-\alpha_{2}},\,....\,,E_{\alpha_{n}},E_{\psi}\right\} \subset g_{h-1}.
\]

A Lie algebra can have more than one different Cartan subalgebra.
We can obtain a new Cartan subalgebra in following way: we can define
the degree one generator $E=\sum_{i=0}^{r}\sqrt{m_{i}}E_{\alpha_{i}}$,
where $\alpha_{0}=-\psi$, $m_{0}\equiv1$ and the other real constants
are $m_{i}=n_{i}\alpha_{i}^{2}/2$ (no sum in the repeated indices).
This generator appears naturally in Affine Toda field theories \cite{Freeman:1991xw,Fring:1991me,Fring:1991gh}.
It is a normal generator, i.e., 
\[
[E,E^{\dagger}]=0,
\]
which means it can be diagonalized. Therefore the generators $E$
and $E^{\dagger}$ belong to a new Cartan subalgebra $\mathfrak{h'}$\cite{konstant1959}
generated by orthogonal generators $h_{1},h_{2},\ldots,h_{r}$, which
are related by a similarity transformation to the generators $H_{1},H_{2},...,H_{r}$
of the Cartan subalgebra $\mathfrak{h}$, that is,
\[
h_{i}=PH_{i}P^{-1},\,\,\,\,\,i=1,2,\,...\,,r,\,\,\,\,\,P\in G.
\]
Remembering that for arbitrary integers $p_{i}$ the group elements
$\exp\left(2\pi ip_{i}\lambda_{i}^{\vee}\cdot H\right)$ belong to
the center of the group $\mathcal{Z}(G)$, then,\cite{Kneipp:1994sz}
\begin{equation}
\exp\left(2\pi ip_{i}\lambda_{i}^{\vee}\cdot h\right)=\exp\left(2\pi ip_{i}\lambda_{i}^{\vee}\cdot H\right)\in\mathcal{Z}(G),\label{eq:center}
\end{equation}
where we used the fact that since $\exp\left(2\pi ip_{i}\lambda_{i}^{\vee}\cdot H\right)$
lies in the center of the group, it commutes with $P$. 

In particular for the algebra $\mathfrak{su}(3)$ the grading is 
\[
\left\{ H_{1},H_{2}\right\} \subset g_{0},
\]
\[
\left\{ E_{\alpha_{1}},E_{\alpha_{2}},E_{-\psi},E\right\} \subset g_{1},
\]
\[
\left\{ E_{-\alpha_{1}},E_{-\alpha_{2}},E_{\psi},E^{\dagger}\right\} \subset g_{2}.
\]
The generators $E$ and $E^{\dagger}$ are given by

\begin{equation}
E=E_{\alpha_{1}}+E_{\alpha_{2}}+E_{-(\alpha_{1}+\alpha_{2})},\quad E^{\dagger}=E_{-\alpha_{1}}+E_{-\alpha_{2}}+E_{(\alpha_{1}+\alpha_{2})},\label{eq:conventions0}
\end{equation}
or in terms of $h_{i}$,
\begin{equation}
E=\sqrt{3}\left(e^{-i\pi/6}\lambda_{1}+e^{i\pi/2}\lambda_{2}\right)\cdot h,\quad E^{\dagger}=\sqrt{3}\left(e^{i\pi/6}\lambda_{1}+e^{-i\pi/2}\lambda_{2}\right)\cdot h.\label{eq:conventions1}
\end{equation}
It is convenient to define a new basis where the other six $\mathfrak{su}(3)$
generators are given by 
\begin{equation}
X_{0}^{1}\equiv\frac{1}{\sqrt{3}}(\alpha_{1}-\alpha_{2})\cdot H,\quad X_{0}^{2}\equiv\psi\cdot H,\label{eq:conventions2}
\end{equation}
\begin{equation}
X_{1}^{1}\equiv\frac{1}{\sqrt{3}}(E_{\alpha_{1}}+E_{\alpha_{2}}-2E_{-\psi}),\quad X_{1}^{2}\equiv-(E_{\alpha_{1}}-E_{\alpha_{2}}),\label{eq:conventions3}
\end{equation}
\begin{equation}
X_{2}^{1}\equiv(X_{1}^{1})^{\dagger},\quad X_{2}^{2}\equiv(X_{1}^{2})^{\dagger},\label{eq:conventions4}
\end{equation}
where $X_{M}^{i}$ has degree $M$ and satisfies $(X_{M}^{i})^{\dagger}=X_{[3-M]}^{i}$,
with $[M]$ denoting $M$ modulo $3$. This set of generators has
the normalization in the adjoint representation
\[
Tr(X_{M}^{i}X_{N}^{j\dagger})=2\delta_{ij}\delta_{MN},\quad Tr(EE^{\dagger})=3,\quad Tr(EX_{M}^{i})=0,
\]
and the commutation relations
\[
[E,X_{M}^{i}]=\epsilon_{ij}\sqrt{3}X_{[M+1]}^{j},\quad[E^{\dagger},X_{M}^{i}]=-\epsilon_{ij}\sqrt{3}X_{[M-1]}^{j}.
\]
From \eqref{eq:conventions4} and the fact that $X_{0}^{i\dagger}=X_{0}^{i}$,
the set $\left\{ X_{M}^{i}\right\} $ have only seven non vanishing
independent commutation relations, which read
\[
[X_{0}^{1},X_{1}^{1}]=-X_{1}^{2},\quad[X_{0}^{1},X_{1}^{2}]=-X_{1}^{1}-\frac{2}{\sqrt{3}}E,\quad[X_{0}^{2},X_{1}^{1}]=-X_{1}^{1}+\frac{2}{\sqrt{3}}E,\quad[X_{0}^{2},X_{1}^{2}]=X_{1}^{2},
\]
\[
[X_{1}^{1},X_{1}^{2}]=\frac{2}{\sqrt{3}}E^{\dagger},\quad[X_{1}^{1},X_{2}^{1}]=-X_{0}^{2},\quad[X_{1}^{1},X_{2}^{2}]=-X_{0}^{1}.
\]
Note that by calculating $Tr(E[X_{M}^{i},X_{N}^{j}])$ and $Tr(E^{\dagger}[X_{M}^{i},X_{N}^{j}])$
one can write the above commutation relations in a more compact form
\[
[X_{M}^{i},X_{N}^{j}]=C_{MN}^{ijk}X_{[M+N]}^{k}+\frac{2}{\sqrt{3}}\epsilon_{ij}(\delta_{[M+N+1],0}E^{\dagger}-\delta_{[M+N-1],0}E),
\]
where $C_{MN}^{ijk}=0,\pm1$, according to the above commutation relations.
With this new basis it turned out to be easier to compute the commutators
needed to obtain the asymptotic fields.

\section{Vacuum configuration}

A vacuum solution
\begin{equation}
\phi_{1}^{vac}=v\cdot H,\label{eq:vaccum1}
\end{equation}
with $v=v_{i}\lambda_{i}^{\vee}$ having all $v_{i}$ different from
zero, commute only with the generators in the Cartan subalgebra and
therefore it spontaneously breaks the gauge symmetry to the maximal
torus $U(1)^{r}$. The elements belonging to this unbroken group are
then written as $\exp\left(ic_{j}\lambda_{j}^{\vee}\cdot H\right)$
with $c_{j}$ being real parameters. We can further consider another
scalar field vacuum
\begin{equation}
\phi_{2}^{vac}=\sum_{j=1}^{r}b_{j}E_{\alpha_{j}},\label{eq:vaccum2}
\end{equation}
with all $b_{j}\neq0$, and by Baker-Campbell-Hausdorff (BCH) formula
one can show that the only elements $\exp\left(ic_{j}\lambda_{j}^{\vee}\cdot H\right)$
leaving this vacuum invariant are
\[
\exp\left(2\pi i\omega\cdot H\right),
\]
with $\omega$ belonging to the co-weight lattice of $G$. Since these
are just the center elements of $G$, we see that this vacuum configuration
produces a spontaneous symmetry breaking pattern 
\begin{equation}
G\xrightarrow{\phi_{1}}U(1)^{r}\xrightarrow{\phi_{2}}\mathcal{Z}(G).\label{eq:symmetry breaking pattern}
\end{equation}

Let us consider the same potential discussed in \cite{Kneipp:2003ue,Kneipp:2007fg},
\[
V=\frac{1}{2}d_{a}^{2},\quad d_{a}=\frac{e}{2}\left(\sum_{s=1}^{2}[\phi_{s}^{\dagger},\phi_{s}]-m\frac{\phi_{1}+\phi_{1}^{\dagger}}{2}\right),
\]
where $m$ is a real mass parameter, which accept vacuum solutions
of the form (\ref{eq:vaccum1}) and (\ref{eq:vaccum2}). In the case
of $G=SU(3)$ these vacuum solutions can be written as
\[
\phi_{1}^{vac}=a_{1}\psi\cdot H=a_{1}X_{0}^{2},
\]
where $\psi=\alpha_{1}+\alpha_{2}=\lambda_{1}+\lambda_{2}$ is the
highest root of $SU(3)$, and
\[
\phi_{2}^{vac}=a_{2}\left(E_{\alpha_{1}}+E_{\alpha_{2}}\right)=\frac{a_{2}}{3}\left(\sqrt{3}X_{1}^{1}+2E\right),
\]
where $a_{1}$ and $a_{2}$ are real constants. This vacuum configuration
spontaneously breaks the gauge symmetry in the pattern 
\[
SU(3)\xrightarrow{\phi_{1}}U(1)\times U(1)\xrightarrow{\phi_{2}}\mathbb{Z}_{3},
\]
giving rise to a multiply connected vacuum manifold which allows the
existence of $\mathbb{Z}_{3}$ strings.

\section{Asymptotic $\mathbb{Z}_{3}$ string solutions}

For a theory with gauge group $G$ broken to its center $\mathcal{Z}(G)$,
the energy per unit length or string tension of a static topological
non-Abelian string, considering $W_{0}=0=W_{3}$, is given by
\[
T=\int d^{2}x\left[\frac{1}{4}G_{ij}^{a}G_{ij}^{a}+\frac{1}{2}|D_{i}\phi_{s}|^{2}+V(\phi_{s})\right],
\]
where $i=1,2$ denotes directions perpendicular to the string. In
order to the string tension be finite, the asymptotic form of the
fields must be related to the vacuum configuration by a gauge transformation,
\begin{equation}
\begin{aligned}W_{i}(\varphi) & =\frac{i}{e}\left(\partial_{i}g(\varphi)\right)g(\varphi)^{-1},\\
\phi_{s}(\varphi) & =g(\varphi)\phi_{s}^{vac}g(\varphi)^{-1}.
\end{aligned}
\label{eq:general asymptotic fields}
\end{equation}
 In order to be single valued configurations the group element $g(\varphi)$
satisfies
\begin{equation}
g(\varphi+2\pi)g(\varphi)^{-1}\in\mathcal{Z}(G).\label{eq:intro2}
\end{equation}
Notice that by assuming $g(0)$ as the identity, then the above condition
leads to $g(2\pi)\in\mathcal{Z}(G).$ We can consider that 
\begin{equation}
g(\varphi)=\exp(i\varphi M).\label{eq:intro3}
\end{equation}
Then $\exp(2\pi iM)\in\mathcal{Z}(G)$ and $M$ must be diagonalizable
which implies that $M$ must be a normal generator, that is, $\left[M,M^{\dagger}\right]=0$.
In order to fulfill this condition, one can consider that $M=\omega.H$
is a linear combination of Cartan generators. Then, the vector $\omega$
is in the co-weight lattice of the gauge group to guarantee that $\exp(2\pi i\omega\cdot H)\in\mathcal{Z}(G)$,
as was considered in \cite{Kneipp:2003ue,Kneipp:2007fg}. As it can
be seen from equations (\ref{eq:general asymptotic fields}) and (\ref{eq:intro3}),
the asymptotic gauge field is in the Cartan subalgebra of $\mathfrak{g}$.
However, this is not the only possible choice for $M$. From Eq. (\ref{eq:center})
we can see that we can consider $M=p_{i}\lambda_{i}^{\vee}\cdot h,$
$p_{i}\in\mathbb{Z}$. For simplicity we shall consider only the theory
with gauge group $G=SU(3)$. In this case we shall have
\begin{equation}
g(\varphi)=\exp\left[i\varphi(p_{1}\lambda_{1}+p_{2}\lambda_{2})\cdot h\right],\,\,\,\,\,\,p_{1},p_{2}\in\mathbb{Z},\label{eq:asymptotic1}
\end{equation}
recalling that the algebra $\mathfrak{su}(3)$ is simply laced so
that $\lambda_{i}^{\vee}=\lambda_{i}$. Solving (\ref{eq:conventions1})
for $\lambda_{i}\cdot h$ we find 
\begin{equation}
M_{1}\equiv\lambda_{1}\cdot h=\frac{E+E^{\dagger}}{3},\quad M_{2}\equiv\lambda_{2}\cdot h=\frac{e^{-i\pi/3}E+e^{i\pi/3}E^{\dagger}}{3}.\label{eq:asymptotic2}
\end{equation}
It can be checked explicitly that $g(2\pi)$ belongs indeed to the
center of $SU(3)$. In effect, in the 3 dimension representation
\[
E=\left(\begin{array}{ccc}
0 & 1 & 0\\
0 & 0 & 1\\
1 & 0 & 0
\end{array}\right),
\]
and a direct calculation shows that the powers of $M_{1}$ are given
by 
\[
M_{1}^{k}=\frac{J_{k}(3M_{1}+\mathds{1})+(-1)^{k}\mathds{1}}{3^{k}},\quad k\geq0,
\]
where
\[
J_{k}=\frac{2^{k}-(-1)^{k}}{3},\quad k\geq0,
\]
are the Jacobsthal Numbers. Then it is easy to compute
\[
\exp\left(2\pi inM_{1}\right)=\exp\left(\frac{4\pi in}{3}\right)\mathds{1}\in\mathcal{Z}(SU(3)).
\]
In a similar way,
\[
M_{2}^{k}=\frac{(-1)^{k}J_{k}(\mathds{1}-3M_{2})+\mathds{1}}{3^{k}},\quad k\geq0,
\]
and

\[
\exp\left(2\pi inM_{2}\right)=\exp\left(\frac{2\pi in}{3}\right)\mathds{1}\in\mathcal{Z}(SU(3)).
\]
Therefore the group elements given by (\ref{eq:asymptotic1}) are
in the center of $SU(3)$.

For simplicity we shall adopt the notation
\[
M_{\lambda}=\frac{\lambda E+\lambda^{-1}E^{\dagger}}{3},
\]
where $\lambda=1$ or $\lambda=\exp(-i\pi/3)$. Let us consider the
group element $g(\varphi)=\exp(i\varphi nM_{\lambda}),\,\,\,\,n\in\mathbb{Z}$.
Then the asymptotic gauge field reads
\[
W_{i}(\varphi)=-\frac{\epsilon_{ij}x^{j}}{e\rho^{2}}nM_{\lambda},
\]
or in polar coordinates
\[
W_{\rho}(\varphi)=0,\quad W_{\varphi}(\varphi)=-\frac{1}{e}nM_{\lambda}.
\]

Let us define the generators
\[
\begin{aligned}S_{\lambda}^{1} & =\frac{\sqrt{3}}{3}\left(-\lambda X_{1}^{1}+\lambda^{-1}X_{2}^{1}\right),\\
S_{\lambda}^{2} & =\frac{1}{3}\left(2X_{0}^{2}-\lambda^{-2}X_{1}^{2}-\lambda^{2}X_{2}^{2}\right),
\end{aligned}
\]
which satisfy
\[
[M_{\lambda},S_{\lambda}^{1}]=S_{\lambda}^{2},\quad[M_{\lambda},S_{\lambda}^{2}]=S_{\lambda}^{1}.
\]
Using these relations and the BCH formula we obtain the asymptotic
form of $\phi_{1}$ as
\begin{equation}
\phi_{1}(\varphi)=\phi_{1}^{vac}+a_{1}\left\{ i\sin(n\varphi)S_{\gamma}^{1}+\left[\cos(n\varphi)-1\right]S_{\gamma}^{2}\right\} .\label{eq:asymptotic3}
\end{equation}
Similarly we define
\[
\begin{aligned}T_{\lambda}^{1} & =\frac{\sqrt{3}}{9}\left(2X_{1}^{1}-\lambda^{2}X_{0}^{1}-\lambda^{-2}X_{2}^{1}\right),\\
T_{\lambda}^{2} & =\frac{1}{3}\left(\lambda X_{2}^{2}-\lambda^{-1}X_{0}^{2}\right),
\end{aligned}
\]
such that these generators obey
\[
[M_{\lambda},T_{\lambda}^{1}]=T_{\lambda}^{2},\quad[M_{\lambda},T_{\lambda}^{2}]=T_{\lambda}^{1},
\]
and then we obtain
\begin{equation}
\phi_{2}(\varphi)=\phi_{2}^{vac}+a_{2}\left\{ i\sin(n\varphi)T_{\lambda}^{2}+\left[\cos(n\varphi)-1\right]T_{\lambda}^{1}\right\} .\label{eq:asymptotic4}
\end{equation}

Considering that 
\[
W_{i}(\varphi,\rho)=-g(\rho)\frac{\epsilon_{ij}x^{j}}{e\rho^{2}}nM_{\lambda}
\]
where $g(\rho)$ is a radial function with $g(\infty)=1$ and $g(0)=0$,
then the non-vanishing component of the magnetic field has the form
\[
B_{3}(\varphi,\rho)=-G_{12}(\varphi,\rho)=g'(\rho)\frac{nM_{\lambda}}{e\rho}.
\]
Therefore, for these strings, the gauge field and the magnetic field
take value in the direction of the step operators and we will call
them E-strings. Since we are considering the same potential as in
our previous works, this theory also have $\mathbb{Z}_{3}$ string
in the direction of the Cartan subalgebra, which we shall call H-strings.
Due to relation (\ref{eq:center}), E-strings belong to the same three
topological sectors as the H-string. Which one is stable will depend
on the form of the ansatz and energetic considerations which we will
analyse in another work. It is interesting to note that in the symmetry
breaking pattern (\ref{eq:symmetry breaking pattern}), there appear
monopoles in the first breaking, which get confined in the second
breaking \cite{Kneipp:2003ue}. However, since the monopoles have
magnetic flux in the Cartan direction, the will get confined only
by H-strings. Then, in order to have finite energy the E-strings should
not end in monopoles but close into itself.

\bibliographystyle{ieeetr}
\bibliography{MyCollection}

\end{document}